\documentclass[prapplied,twocolumn,
superscriptaddress,showkeys]{revtex4-1}
\usepackage{graphicx}
\usepackage{hyperref}
\usepackage{dcolumn}   
\usepackage[english]{babel}
\usepackage{bm}        
\usepackage{amssymb}   
\usepackage{inputenc}
\usepackage{amsmath}
\usepackage{amssymb}
\usepackage{graphicx}
\usepackage{natbib}
\usepackage{mathrsfs}
\usepackage{color}
\usepackage{upgreek}
\usepackage{wrapfig}
\usepackage{comment}
\usepackage{adjustbox}
\usepackage[normalem]{ulem} 
\hypersetup{breaklinks=true}

\usepackage{etoolbox} 

\usepackage{gensymb}

\footnotetext{L.S. and D.M.L. contributed equally to this paper.}

\begin{document}
\title{Low-Dissipation Nanomechanical Devices from Monocrystalline Silicon Carbide}
\author{Leo Sementilli$^\dagger$}
\affiliation{The Australian Research Council Centre of Excellence for Engineered Quantum Systems, School of Mathematics and Physics, University of Queensland, St. Lucia, Queensland 4072, Australia}
\author{Daniil M. Lukin$^\dagger$}
\affiliation{E. L. Ginzton Laboratory, Stanford University, Stanford, California 94305, USA}
\author{Hope Lee}
\affiliation{E. L. Ginzton Laboratory, Stanford University, Stanford, California 94305, USA}
\author{Joshua Yang}
\affiliation{E. L. Ginzton Laboratory, Stanford University, Stanford, California 94305, USA}
\author{Erick Romero}
\affiliation{The Australian Research Council Centre of Excellence for Engineered Quantum Systems, School of Mathematics and Physics, University of Queensland, St. Lucia, Queensland 4072, Australia}
\author{Jelena Vu$\check{\textnormal{c}}$kovi$\acute{\textnormal{c}}$}
\email{jela@stanford.edu}
\affiliation{E. L. Ginzton Laboratory, Stanford University, Stanford, California 94305, USA}
\author{Warwick P. Bowen}
\affiliation{The Australian Research Council Centre of Excellence for Engineered Quantum Systems, School of Mathematics and Physics, University of Queensland, St. Lucia, Queensland 4072, Australia}
\affiliation{The Australian Research Council Centre of Excellence in Quantum Biotechnology, School of Mathematics and Physics, University of Queensland, St. Lucia, Queensland 4072, Australia}


\maketitle

\section{abstract}

The applications of nanomechanical resonators range from biomolecule mass sensing to hybrid quantum interfaces. Their performance is often limited by internal material damping, which can be greatly reduced by using crystalline materials. Crystalline silicon carbide is appealing due to its exquisite mechanical, electrical and optical properties, but has suffered from high internal damping due to material defects. Here we resolve this by developing nanomechanical resonators fabricated from bulk monocrystalline 4H-silicon carbide. This allows us to achieve damping as low as 2.7~mHz, more than an order-of-magnitude lower than any previous crystalline silicon carbide resonator and corresponding to a quality factor as high as 20 million at room temperature. The volumetric dissipation of our devices reaches the material limit for silicon carbide for the first time. This provides a path to greatly increase the performance of silicon carbide nanomechanical resonators.

\section{Introduction}

Nanomechanical resonators have many applications, from nanoscale probes in biological environments~\cite{chien_single-molecule_2018,roslon_probing_2022}, to radio and microwave frequency timing~\cite{massel_microwave_2011,nguyen_mems_2007}, and on-chip navigation and position awareness~\cite{krause_high-resolution_2012,fan_graphene_2019}. The mechanical dissipation rate is a key figure of merit~\cite{sementilli_nanomechanical_2022}. It is ultimately determined by material friction and dictates the sensitivity limits of nanomechanical sensors, accuracy of nanomechanical timing, and sharpness of nanomechanical filters. In principle pure crystalline materials have lower material dissipation than amorphous counterparts because of their ordered atomic lattice. However, defects introduced during material growth, processing, and nanofabrication often result in dissipation rates higher than those of amorphous materials~\cite{romero_engineering_2020,buckle_nanomechanical_2020}. 

Crystalline silicon carbide (SiC) is an important material for nanomechanics, as it possesses many attractive mechanical, electronic and optical properties. It has high material yield strength, can be fabricated into high stress thin films~\cite{petersen_silicon_1982,romero_engineering_2020,severino_3c-sic_2011,kermany_microresonators_2014}, has high thermal conductivity and wide electronic bandgap~\cite{perret_power_2009}, and hosts color centers used for quantum photonics~\cite{lukin_two-emitter_2023,janitz_cavity_2020}. Furthermore, it has excellent photonic properties~\cite{guidry_quantum_2022,lukin_4h-silicon-carbide--insulator_2020,castelletto_silicon_2020}, can be mass manufactured in industrial settings, and sold as an affordable semiconductor. However, to date, all crystalline silicon carbide nanomechanical resonators have had material dissipation orders-of-magnitude higher than the predicted volumetric limit~\cite{kermany_microresonators_2014,romero_engineering_2020,klas_high_2022,huang_quality_2003}.

Here, we develop crystalline silicon carbide nanomechanical devices with ultra-low dissipation, utilizing bulk sublimation grown silicon carbide crystals and a grind-and-polish technique to achieve defect-free thin films. This eliminates the interfacial defect layer that causes dissipation in silicon carbide nanomechanical resonators fabricated directly on silicon substrates~\cite{kermany_microresonators_2014,romero_engineering_2020}. We observe greatly reduced intrinsic damping, achieving dissipation rates as low as 2.7~mHz at room temperature. This is nearly two orders-of-magnitude lower than what has been achieved in bulk crystalline silicon carbide resonators~\cite{hamelin_monocrystalline_2019}, an order-of-magnitude lower than silicon carbide nanomechanical resonators fabricated from heteroepitaxially grown crystals~\cite{romero_engineering_2020}, and a factor of $1.6\times$ better than the best reported in high stress amorphous silicon carbide resonators~\cite{xu_high-strength_2023}. It corresponds to a quality factor as high as 20 million, even with only a few hundred megapascal of tensile stress. Fabrication from bulk silicon carbide allows us to reach volumetric dissipation at the reported material limit~\cite{ashby_overview_1989,schmid_fundamentals_2023}.
 
The low linear dissipation allows us to make the first observation of nonlinear dissipation in crystalline silicon carbide nanomechanical resonators. We find that this is lower than other materials such as amorphous silicon nitride. This is important for applications where nonlinear effects constrain performance, such as mass sensing~\cite{ekinci_ultimate_2004,yang_zeptogram-scale_2006}, and nanomechanical computing~\cite{mauranyapin_tunneling_2021,romero_acoustically_2024}. 

\begin{figure*}[ht]
\centering
\includegraphics[width=\textwidth]{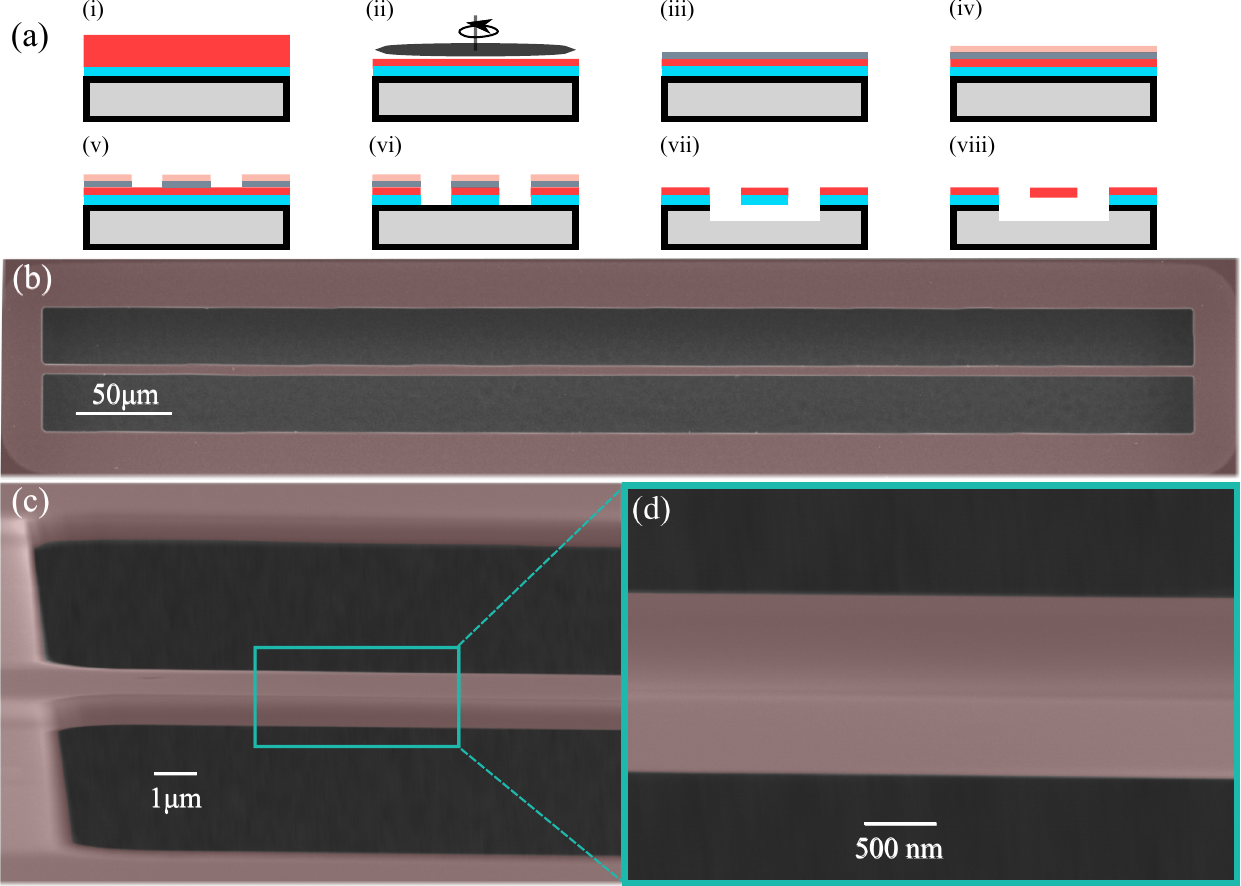}
\caption{(a) Fabrication process. Colors correspond to red-SiC, light blue-SiO2, dark gray-aluminum, salmon-resist, light gray-silicon substrate.(b) SEM image of a nanostring resonator made of 4H-SiC. (c)  SEM image of the sidewall and clamping point of a nanostring resonator. (d) Higher magnification SEM image of nanostring sidewall from (c). Minimal surface roughness is observed in SEM images taken with 10~kV accelerating voltage and 17~k magnification.}\label{fig:StanSEMs}
\end{figure*}

 At the material damping limit and when strained to near the material yield strength (21~GPa~\cite{petersen_silicon_1982}), crystalline silicon carbide resonators could allow quality factors exceeding 10 billion at room temperature. This would surpass the achievable quality factors of other materials, such as silicon and silicon nitride, which are currently limited by higher intrinsic nanomechanical dissipation and lower yield strengths~\cite{sementilli_nanomechanical_2022,villanueva_evidence_2014,yasumura_quality_2000,petersen_silicon_1982,norte_mechanical_2016}. Realizing such extreme nanomechanical quality factors at room temperature, while challenging, may shed light on new dissipation mechanisms and enable fundamentally new applications such as nanomechanical tests of spontaneous wavefunction collapse~\cite{forstner_nanomechanical_2020}, sensing of dark matter~\cite{carney_mechanical_2021}, and room-temperature quantum optomechanics~\cite{guo_feedback_2019,norte_mechanical_2016}.

\section{Device Fabrication}

The devices studied here are fabricated using a process consisting of thin film preparation, metal deposition, electron beam patterning, reactive ion etching, and dry selective release following the process outlined in Ref.~\cite{lukin_4h-silicon-carbide--insulator_2020}~(see Figure~\ref{fig:StanSEMs}(a)). The samples are first derived from bulk crystalline 4H-SiC wafers, which are then thermally bonded onto silicon carrier wafers via a thin (160 nm) bonding silicon oxide layer. The bonding procedure requires annealing at high temperatures ($900\degree \textnormal{C}$)~\cite{lukin_4h-silicon-carbide--insulator_2020}, which introduces stress into the SiC layer upon thermalization to room temperature. The magnitude of this stress is determined by the difference in thermal expansion coefficients between silicon and silicon carbide. The bonded SiC film is then thinned to sub-micron thickness using the grind-and-polish technique~\cite{lukin_4h-silicon-carbide--insulator_2020}. This technique, developed in recent years, has been used to enable low-loss integrated photonics in 4H-SiC and diamond~\cite{lukin_4h-silicon-carbide--insulator_2020,hausmann_integrated_2012}, as well as, most recently, on-chip titanium-sapphire lasers~\cite{yang_titaniumsapphire--insulator_2024}.  The thinning process results in thickness nonuniformity of approximately 10-20~nm per millimeter of resonator length across our sample, which we determine using optical thin film profiling techniques. To account for this, mechanical resonators are selectively patterned within regions of largest mapped uniformity. The mean device thickness is then found across each resonators footprint and used for analysis purposes. 

After the preparation of the crystalline thin film, aluminum is evaporated on the SiC layer to act as a hard mask for etching. Following this, device geometries are initially realized using electron beam lithography, then formed using reactive ion etching of the aluminum, SiC, and silica layers. The aluminum is then stripped chemically, and the devices are undercut using XeF$_2$ dry etching. This leaves 4H-SiC structures suspended with thermal silica still adhered to the bottom interface of the devices. The remaining thermal silica is then removed using vapor HF, resulting in freestanding structures that are purely 4H-SiC. 

We focus our experiments on tensile stressed high-aspect-ratio nanostring resonators, which inherently possess large dissipation dilution factors that allow quality factors far above material loss limits~\cite{kermany_factors_2016,schmid_damping_2011,fedorov_generalized_2019} and are amenable to analytical modelling~\cite{sementilli_nanomechanical_2022,schmid_fundamentals_2023}. SEM images of a completed nanostring are presented in Figures~\ref{fig:StanSEMs}(b-d). Within these images, both the device's top surfaces and sidewalls appear relatively smooth. To quantify surface roughness, we perform a spatial autocorrelation on a section of the resonator's vertical sidewall in Figure~\ref{fig:StanSEMs}(d). From this, we observe roughness correlations on the length scale of 6~nanometers, which both sets a limit to surface feature sizes we can detect with SEM and speaks to the extent of observable roughness in our device sidewalls.

\section{Results}\label{sec:results}

We characterize the dissipation and quality factors of the nanomechanical resonators using ringdown measurements in an optical heterodyne detection setup~\cite{romero_engineering_2020,kermany_microresonators_2014}. We perform ringdown measurements for the first three transverse modes of twenty high-aspect-ratio nanostring devices with lengths ($L$) of 3.1~mm and thicknesses ($h$) between 110-135~nm.

\begin{figure}[ht]
\centering
\includegraphics[width=0.48\textwidth]{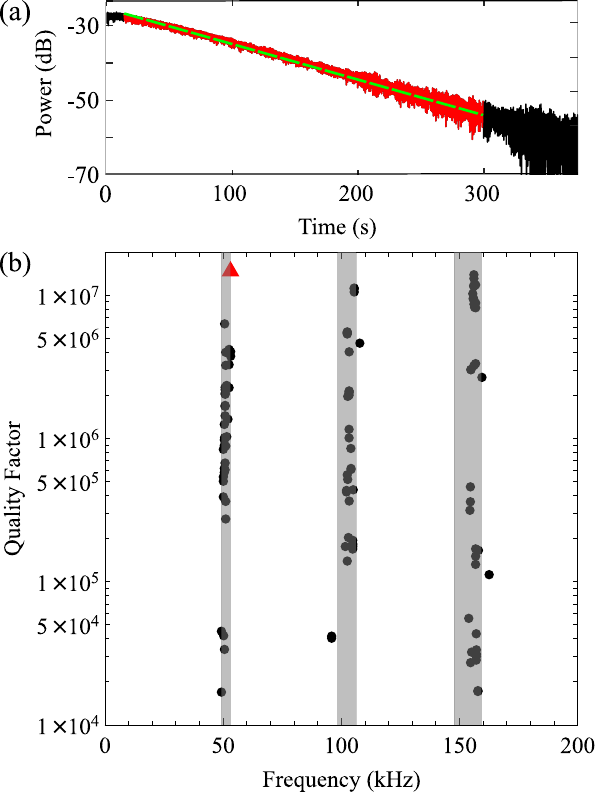}
\caption{(a) Ringdown measurement of a device with a quality factor exceeding $10^{7}$ at 53 kHz. (b) $\textnormal{Q}\cdot f$ map for the first three transverse modes of 20 high aspect-ratio nanostrings. The shaded gray regions represent the expected frequency range based on device dimensions and
tensile stress. The red colored point represents the measurement shown in (a).}\label{fig:QvsfDDStrings}
\end{figure}

An example experimental ringdown measurement on a nanostring with a resonance frequency of $53~\textnormal{kHz}$ is shown in Figure~\ref{fig:QvsfDDStrings}(a). It provides a quality factor of $Q=1.5 \cdot 10^{7}$. The measured quality factors for the first three transverse mechanical modes of all twenty devices are plotted in Figure~\ref{fig:QvsfDDStrings}(b) as a function of resonance frequency. Quality factors exceeding $10^7$ are observed for all three modes. The gray shaded regions represent the expected eigenfrequency range for the first three transverse mechanical modes ($n=1,2,3$) based on the analytical expression~\cite{schmid_fundamentals_2023} 

\begin{equation}\label{eq:eigenfreq}
f=\frac{{n}}{2L}\sqrt{\frac{\sigma}{\rho}},
\end{equation}
 
\noindent where $\sigma$ is the stress and $\rho=3.2~ \textnormal{g/cm}^3$~\cite{kimoto_fundamentals_2014} is the material density. We find that the stress of the devices range from $\sigma=290~\textnormal{MPa}-335~\textnormal{MPa}$ from the measured minimum and maximum fundamental resonances frequencies. Extrapolating this range to the second and third resonances, we find good agreement between theory and experiment (gray bounds in Figure~\ref{fig:QvsfDDStrings}(b)).

The mechanical dissipation rate of each mode can be determined as $\Gamma / 2\pi=f/Q$. We find a minimum mechanical dissipation rate of 2.7~mHz, achieved for the fundamental transverse mode of a high-aspect-ratio nanostring (observed from nonlinear characterization later in Section~\ref{sec:Nonlinear}). This is the lowest dissipation rate reported for any silicon carbide mechanical resonator to-date. It is more than an order-of-magnitude better than the best reported previously in crystalline 3C-SiC nanostrings~\cite{kermany_factors_2016}. The lower
dissipation realized here can be largely attributed to the thin film preparation technique allowing high quality crystalline resonators, but also in-part to the larger device aspect ratio realized in this work~\cite{kermany_factors_2016}. Despite the significant reduction in the dissipation of crystalline SiC resonators, our results are only marginally better ($1.66\times$ - see Table~\ref{table:1}) than the lowest dissipation rate achieved in amorphous SiC resonators~\cite{xu_high-strength_2023}. However, to achieve low dissipation Ref.~\cite{xu_high-strength_2023} employed soft-clamping techniques and higher stress than our work. Applied into crystalline SiC resonators, these techniques have potential to further reduce dissipation beyond the values we report.

While 4H-SiC resonators provide exceptionally low dissipation, the current fabrication procedure results in five times less tensile stress than previous silicon carbide devices in literature~\cite{kermany_factors_2016}. Hence, one would expect lower quality factors and $Q \cdot f$ values than previous demonstrations, since both dissipation dilution factors and frequency increase with tensile stress~\cite{sementilli_nanomechanical_2022,schmid_fundamentals_2023}. Despite this, our highest measured quality factor of 20 million at room temperature is nearly an order of magnitude higher than the best reported for crystalline SiC nanostring and trampoline nanomechanical resonators in the literature~\cite{romero_engineering_2020,kermany_factors_2016,klas_high_2022}. Furthermore, our best $Q\cdot f$ product of $1\cdot 10^{12}$ in these devices exceeds the highest reported $Q\cdot f$ product in high stress ($\sigma=1.5~\textnormal{GPa}$) crystalline 3C-SiC nanostrings~\cite{kermany_microresonators_2014}. The high $Q \cdot f$ product despite low device stress demonstrated here challenges the standard approach where high stress is utilized to increase both resonance frequency and quality factors~\cite{schmid_fundamentals_2023,sementilli_nanomechanical_2022}.

To form an understanding of intrinsic dissipation mechanisms in 4H-SiC we combine the quality factor measurements of high-aspect-ratio nanostrings, together with a suite of ringdown measurements from eighteen other cantilever resonators and thirty low-aspect-ratio nanostrings (see supporting information). The additional resonator geometries are needed to separate the contributions of surface and volumetric effects towards the total intrinsic material dissipation. Cantilever resonators possess no tensile stress and low radiation loss (see supporting information), therefore their measured quality factors well approximate the intrinsic quality factor, $Q_{\textnormal{int}}$, of the material~\cite{schmid_fundamentals_2023,villanueva_evidence_2014,romero_engineering_2020}. String resonators possess tensile stress and dissipation dilution, as well as non-negligible radiation loss. Therefore, the measured quality factors of string resonators do not exactly reflect the intrinsic quality factor. In order to calculate the intrinsic quality factor of each high and low-aspect-ratio nanostring, we determine its dissipation dilution factor using the analytical expression~\cite{schmid_fundamentals_2023,sadeghi_influence_2019}
\begin{equation}
    D \approx \Bigg[\frac{(n\pi)^2}{12}\frac{E}{\sigma}\Big(\frac{h}{L}\Big)^2 +\frac{1}{\sqrt{3}}\sqrt{\frac{E}{\sigma}}\Big(\frac{h}{L}\Big)\Bigg]^{-1} 
    \label{eq:DDApproxCorrect},
\end{equation}
where $E=440~\textnormal{GPa}$ is the Young's Modulus of 4H-SiC~\cite{islam_structural_2015}. Using Equation~\ref{eq:DDApproxCorrect}, and the relationship between intrinsic quality factors and dissipation diluted quality factors ($Q_{\textnormal{D}} = Q_{\textnormal{int}} \times D$), we extract an intrinsic quality factor for nanostring resonators of

\begin{equation}
Q_{\textnormal{int}}=(D \times (Q_{\textnormal{D}}^{-1}-Q_{\textnormal{rad}}^{-1}))^{-1},
\end{equation}

\noindent where $Q_{\textnormal{rad}}$ is the radiation loss limited quality factor (see supporting information)~\cite{romero_engineering_2020}. We include this loss mechanism for all high- and low-aspect-ratio nanostrings, using each device dimension and intrinsic stress inferred from the fundamental transverse eigenfrequency.

\begin{figure}[h]
\centering
\includegraphics[width=0.48\textwidth]{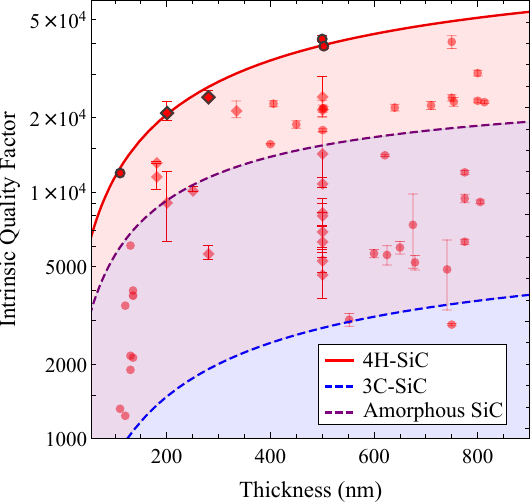}
\caption{Cumulative extracted intrinsic quality factors for all devices, each nanostring is represented by a red circular point and each cantilever is represented by a red diamond. We fit the upper bounds of our extracted intrinsic quality factors (5 red points with black borders) as a function of thickness using Equation~\ref{eq:QintModel}, and the fitting parameter $\alpha$ accounting for radiation loss in nanostring resonators, and plot it using a red line. We add $\textnormal{Q}_\textnormal{int}$ models for amorphous (dashed purple) and crystalline 3C-SiC (dashed blue) for comparison~\cite{xu_high-strength_2023,romero_engineering_2020}.}\label{fig:QvshAllMats}
\end{figure}

We plot the extracted intrinsic quality factors for all high- and low-aspect-ratio nanostrings alongside our cantilever quality factors in Figure~\ref{fig:QvshAllMats}. More than 20 devices have intrinsic quality factors above $10^4$, including at least one device from each of the three sample sets. The maximum intrinsic quality factor from the data is $4.2 \cdot 10^4$ in a 500~nm thick nanostring resonator. No significant differences are observed between the intrinsic quality factors of cantilever and string geometries. This is expected as intrinsic quality factors are primarily determined by a resonators surface-to-volume ratio rather than resonator type~\cite{schmid_fundamentals_2023}. This indicates that the dissipation dilution and radiation loss models of nanostring resonators are appropriate for all nanostrings measured. The intrinsic quality factor increases with device thickness. This is anticipated since surface losses become less important as the surface area to volume ratio of the devices decreases. Similar dependence has been commonly observed in silicon nitride~\cite{villanueva_evidence_2014} and other nanomechanical resonators~\cite{jinling_yang_energy_2002,romero_engineering_2020,yasumura_quality_2000}.

To determine the upper limit of the intrinsic quality factor, a least-square fit is performed among the five red data points with black boundaries in Figure~\ref{fig:QvshAllMats}. We choose these five points because they represent the highest measured intrinsic quality factors at different device thicknesses. They therefore provide information about the highest intrinsic quality factors achieved for the 4H-SiC nanomechanical resonators in this study. The least-square fit follows the process from Ref.~\cite{romero_engineering_2020}, using the standard nanomechanical volume and surface dissipation model given by~\cite{schmid_fundamentals_2023,villanueva_evidence_2014}

\begin{equation}\label{eq:QintModel}
Q_{\textnormal{int}}(h)=(Q_{\textnormal{vol}}^{-1}+(Q_{\textnormal{surf}} \cdot h)^{-1})^{-1},
\end{equation}

\noindent with an additional fitting parameter $\alpha$, which accounts for radiation loss of nanostring resonators (see supplemental information). It is plotted as the red line in Figure~\ref{fig:QvshAllMats} and yields $Q_{\textnormal{vol}}=1.5\cdot 10^5$, $Q_{\textnormal{surf}}=11.5 \cdot 10^{10}~\textnormal{m}^{-1} \cdot h$, and $\alpha=317$. From it, we find that surface loss becomes the dominating intrinsic loss mechanism at roughly 1300~nm thickness. 

The highest observed volumetric quality factor of our resonators is consistent with the theoretical material limit ($1\cdot10^5$) based on the material loss tangent of silicon carbide~\cite{ashby_overview_1989,schmid_fundamentals_2023}. We measure five devices at this material limit, representing $7.5\%$ of our total devices measured. Across three different resonator geometries encompassing 68 total resonators, we find the likelihood that devices reach the material limit decreases as device footprint increases (see supporting information). This is expected under the assumption that the thin films have a uniform density of defects per area and therefore larger devices will encounter more defects. We hypothesis that only partial regions of the thin films are effectively defect-free  due to the introduction of local crystalline imperfections during thin film preparation. 

\section{Observation and Quantification of Nonlinear Dissipation}\label{sec:Nonlinear}

\begin{figure}
\centering
\includegraphics[width=.48\textwidth]{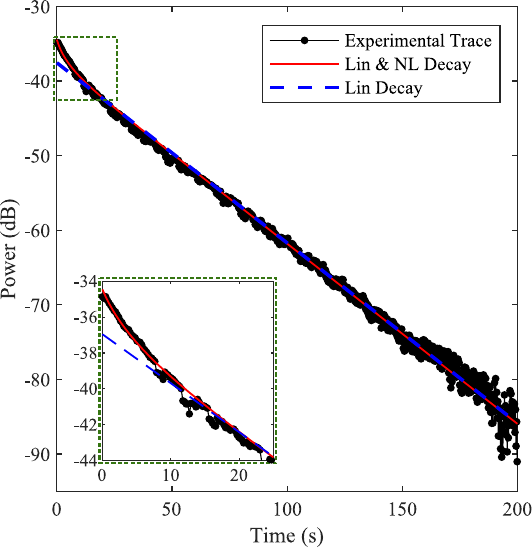}
\caption{Ringdown measurement and analytical fits of a strongly driven, high-Q nanostring resonator. The black trace represents experimental data for the fundamental transverse mode of a high-aspect-ratio nanostring. The red trace represents the fit using a linear and nonlinear decay term~\cite{catalini_modeling_2021}. The blue dashed trace represents the decay of a damped harmonic oscillator, used to fit the
ringdowns of linear nanomechanical resonators.}\label{fig:NonlinearRingdown}
\end{figure}

For applications such as resonant mass sensing~\cite{ekinci_ultimate_2004} and nanomechanical computing~\cite{romero_acoustically_2024}, where the resonator is driven to high amplitudes, it is important to quantify not just linear dissipation but also nonlinear dissipation. The nonlinear dissipation has been characterized in high stress amorphous silicon nitride~\cite{catalini_modeling_2021}, but has yet to be determined in crystalline silicon carbide resonators. To determine it, we strongly drive our high-aspect-ratio nanostrings, and conduct ringdown measurements as shown in Figure~\ref{fig:NonlinearRingdown}.


From this figure, it is apparent the experimental trace deviates from a standard linear decay (blue dashed line) at high amplitudes. To account for this, we include a nonlinear damping term into the ringdown model for dissipation-diluted nanomechanical resonators~\cite{catalini_modeling_2021}. This allows us to extract the linear dissipation rate as well as the nonlinear damping loss parameter. We find the linear mechanical dissipation rate to be 2.7~mHz, a linearly damped mechanical quality factor of $2.0\cdot 10^7$, and a nonlinear damping parameter of $1.1\cdot10^{13}~ \textnormal{s}^{-1}\textnormal{m}^{-2}$. The nonlinear damping parameter is similar to but lower than the lowest that has been experimentally determined in silicon nitride resonators of similar thickness ($\approx 1.5\cdot10^{13}-1\cdot10^{16}~ \textnormal{s}^{-1}\textnormal{m}^{-2}$)~\cite{catalini_modeling_2021}. This suggests
that 4H-SiC resonators may be more linear than those composed of silicon nitride, allowing improved performance. 

\section{Comparison with State-of-the-art}\label{sec:compare}
 
To compare our results to current state-of-the-art devices in silicon carbide, we plot models of the intrinsic quality factor of 3C-SiC and amorphous SiC nanomechanical resonators in Figure~\ref{fig:QvshAllMats} (dashed lines)~\cite{romero_engineering_2020,xu_high-strength_2023}. Using the parameters of these intrinsic quality factor models we find that the 4H polytype has a 23 times higher volumetric quality factor, and 11 times higher surface quality factor compared to 3C-SiC. Interestingly, for the specific case when the interfacial defect-layer is removed from 3C-SiC nanomechanical resonators, its surface loss agrees to within $5\%$ of that found for 4H-SiC here~\cite{romero_engineering_2020}. However, even with the interfacial defect layer removed in 3C-SiC, its volumetric quality factor is still over an order-of-magnitude lower than that of 4H-SiC~\cite{romero_engineering_2020}. This suggests there is an additional cause, in addition to the defect layer, which is partly responsible for the difference in volumetric quality factors between 3C- and 4H-SiC. We speculate this may be related to the cubic versus hexagonal crystalline structure of each respective polytype but requires
further study.

The 4H-SiC nanomechanical resonators from this work also outperform amorphous SiC, with $40\%$ less surface damping and five times less volumetric damping~\cite{xu_high-strength_2023}. While these comparisons are between the upper bounds of the intrinsic quality factor of each material, we measure many resonators with intrinsic quality factors above the limits of both crystalline 3C and amorphous silicon carbide ($98\%$ exceed the upper limit for 3C-SiC and $41\%$ exceed it for amorphous SiC).

Compared to the highest reported values for silicon nitride resonators at room temperature, 4H-SiC possesses a volumetric quality factor five times larger, and a comparable surface quality factor (4H-SiC is $15\%$ greater)~\cite{villanueva_evidence_2014}. Although the dependence of nanomechanical quality factor on device thickness has not been fully characterized for crystalline silicon and diamond, we can compare measured total intrinsic quality factors for cantilever resonators of thicknesses between 100 and 300~nm to 4H-SiC. The intrinsic quality factor of 4H-SiC is roughly $35\%$ higher than the best reported values for single-crystal silicon at room temperature~\cite{tao_single-crystal_2014,yasumura_quality_2000}. It is about three times higher than that reported for polycrystalline diamond, but an order of magnitude less than electronic grade single-crystal diamond~\cite{tao_single-crystal_2014}. 
Applying surface treatments to single-crystal diamond nanomechanical resonators has shown to provide a three to ten times permanent reduction in surface loss~\cite{tao_single-crystal_2014,tao_permanent_2015}. Given the similar crystalline structures of diamond and 4H-SiC, it is conceivable that surface treatments of this kind could also be effective for 4H-SiC nanomechanical resonators.

 Increasing the tensile stress to the material yield strength and using soft-clamped resonator geometries~\cite{beccari_strained_2022,ghadimi_elastic_2018,bereyhi_perimeter_2022}, could allow for quality factors of tens of billions at room temperature in 4H-SiC. Although challenging, if accomplished, this would be comparable to the best cryogenic results using strained silicon nanomechanical resonators at 7 Kelvin~\cite{beccari_strained_2022}, as well as the breathing mode of silicon nanomechanical resonators at millikelvin temperatures~\cite{maccabe_nano-acoustic_2020}.

\begin{table}
    \caption{Comparison of Silicon Carbide nano/bulk mechanical resonators. All measurements are at room temperature. For direct comparison of intrinsic quality factors see Figure~\ref{fig:QvshAllMats}.}
    \resizebox{0.48\textwidth}{!}{
\begin{tabular}{|c|c|c|c|c|}
\hline
     Ref & Frequency & Q ($\times 10^6$) & $\Gamma/2\pi$ (mHz)  & Type\\  \hline
    \cite{hamelin_monocrystalline_2019} & 5.3 MHz & 18 & 290 & 4H-SiCOI Bulk\\  \hline
    \cite{xu_high-strength_2023} & 895 kHz & 198 & 4.5 & Amorphous SiC Nano\\   \hline
    \cite{romero_engineering_2020}  &  211 kHz & 1.74 & 121 & 3C-SiC Nano\\ \hline
    \cite{kermany_microresonators_2014} & 280 kHz & 2.9 & 90 & 3C-SiC Nano\\
 
     \hline
    This Work & 53 kHz & 20 & 2.7  & 4H-SiCOI Nano\\
    \hline
\end{tabular}}\label{table:1} 
\end{table}

\section{Discussion}

While this work reports the lowest dissipation rate achieved to date in a SiC nanomechanical resonator, much lower dissipation has been achieved in silicon nitride resonators due to effective soft-clamping of resonators and high stress~\cite{bereyhi_hierarchical_2022,bereyhi_perimeter_2022,pratt_nanoscale_2023,cupertino_centimeter-scale_2024}. A key difference between our results and this prior work, is that ours is achieved with low stress. This has many practical advantages over highly stressed resonators which are more likely to fail in demanding real-world applications such as inertial and mass sensing~\cite{yang_zeptogram-scale_2006,krause_high-resolution_2012}. Furthermore, it may be possible to reach far lower dissipation levels in crystalline SiC using surface treatments and alternative resonator geometries with higher stress and dissipation dilution such as hierarchical, polygon, and torsional resonators~\cite{bereyhi_hierarchical_2022,bereyhi_perimeter_2022,pratt_nanoscale_2023}. This requires two challenging advances in thin film development. Specifically, the grind-and-polish technique both needs to be successfully extended to films thinner than demonstrated here or in literature~\cite{wang_high-q_2021,song_ultrahigh-q_2019}, and needs to allow for higher tensile stress. Achieving much higher levels of tensile stress will likely require new approaches, similar to those developed for strained silicon-on-insulator wafers~\cite{ghyselen_engineering_2004,beccari_strained_2022}.

Assuming the same resonator geometry and dimensions as Ref~\cite{bereyhi_perimeter_2022}, stressed to half of silicon carbide's yield strength~\cite{petersen_silicon_1982} predicts a diluted quality factor of 18 billion at room temperature. If realized, this would be comparable to the best reported cryogenic nanomechanical resonators~\cite{beccari_strained_2022,maccabe_nano-acoustic_2020}. The benefit of using crystalline SiC over other materials is due to both the lower intrinsic damping and higher material yield strength, which when implemented with a soft-clamped resonator allows dissipation dilution to scale proportionally with strain~\cite{fedorov_generalized_2019,beccari_strained_2022}.

\section{Supporting Information}

Information on measurement setup, additional measurements and analysis, details on radiation loss calculation, additional analysis on rate of material limited devices, FEM modelling of the effect of overhang on dissipation dilution factors

\section{Acknowledgments}

This work was supported by the Australian Research Council Centres of Excellence for Engineered Quantum Systems (EQUS, CE170100009) and Quantum Biotechnology (QUBIC, CE230100021). The authors acknowledge the facilities, and the scientific and technical assistance, of the Australian Microscopy $\&$ Microanalysis Research Facility at the Centre for Microscopy and Microanalysis, The University of Queensland. This research is partially supported by the Commonwealth
of Australia as represented by the Defence Science and Technology Group of the Department of Defence. The work at Stanford was supported by the Vannevar Bush Faculty Fellowship from the Department of Defense.

\clearpage

%

\clearpage
\onecolumngrid

\section*{Supporting Information}

\makeatletter
\setcounter{section}{0}
\setcounter{figure}{0}
\setcounter{equation}{0}
\renewcommand{\theequation}{S-\arabic{equation}}
\renewcommand 
\thesection{S\@arabic\c@section}
\renewcommand\thetable{S\@arabic\c@table}
\renewcommand \thefigure{S\@arabic\c@figure}
\renewcommand{\citenumfont}[1]{S-#1}
\makeatother

\maketitle

\section{Methods}

We characterize the dissipation of the resonators using ringdown measurements in an optical heterodyne detection setup at infrared wavelengths (780 nm) under high vacuum ($\approx 10^{-6}~\textnormal{mbar}$)~\cite{s_romero_engineering_2020,s_kermany_microresonators_2014}. Using this scheme we first measure thermal motion spectra, then employ external piezo driving at the mechanical resonance frequencies to excite the resonators to higher oscillation amplitudes. The external drive piezos are then turned off and the decay of resonator's oscillation amplitude is subsequently monitored with the interferometric detection. An experimental trace of a ringdown measurement is shown in Figure~\ref{fig:QvsLhCanti}(a). This ringdown allows us to determine the quality factor of the mechanical mode, and extract the mechanical dissipation rate. We repeat this process to extract multiple ringdown traces (between 2-5) for each device. 

\section{Additional measurements of Cantilever and String Resonators}

\subsection{Cantilevers}\label{sec:4HCantilevers}
 Uniform cantilever structures are among the simplest mechanical resonators to study because of the lack of tensile stress. This implies that there will be no dissipation dilution and therefore the resonators are subject to only intrinsic and radiation loss (assuming no gas damping). The acoustic radiation loss of a cantilever can be determined using an analytical approximation~\cite{s_schmid_fundamentals_2023}, from which we estimate our cantilevers with the most significant radiation loss to have a radiative quality factor limit of roughly $10^{10}$. As the measured quality factors are several orders-of-magnitude removed from this limit we neglect this contribution in our analysis. Under this assumption, the measured quality factor of the cantilevers are in fact the intrinsic quality factor.

\begin{figure}[h]
\centering
\includegraphics[width=0.48\textwidth]{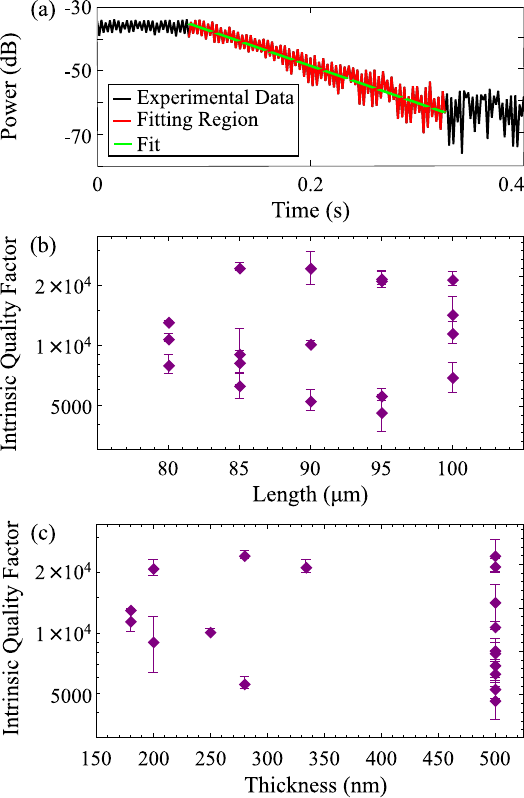}
\caption{(a) Ringdown measurement of one cantilever device. The red fitting region represents free decay of the oscillator, specifically when there is no external drive. In this experimental trace, we fit a quality factor of $2\cdot10^4$ for a fundamental mode at 41.5~kHz. (b) Intrinsic quality factor versus length of cantilever nanomechanical resonators. (c) Intrinsic quality factor versus thickness of cantilever nanomechanical resonators. Each point represents an individual device with error bounded by the standard deviation of many ringdown measurements.}\label{fig:QvsLhCanti}
\end{figure}

Measurements are conducted on 18 different cantilever resonators of constant width ($5~\mu\textnormal{m}$), device lengths between $80-100~\mu \textnormal{m}$, and device thicknesses over a range of $180-500~\textnormal{nm}$. We plot the average intrinsic quality factor measurements as a function of device length and thickness in Figure~\ref{fig:QvsLhCanti}(b-c). Each data point here represents an individual device, with error bounds composed of the standard deviation of measured quality factors for the fundamental mode among multiple traces. We measure a maximum intrinsic quality factor greater than $2\cdot10^{4}$, with the quality factor of many devices exceeding $10^4$. 

\begin{figure*}[h]
\centering
\includegraphics[width=\textwidth]{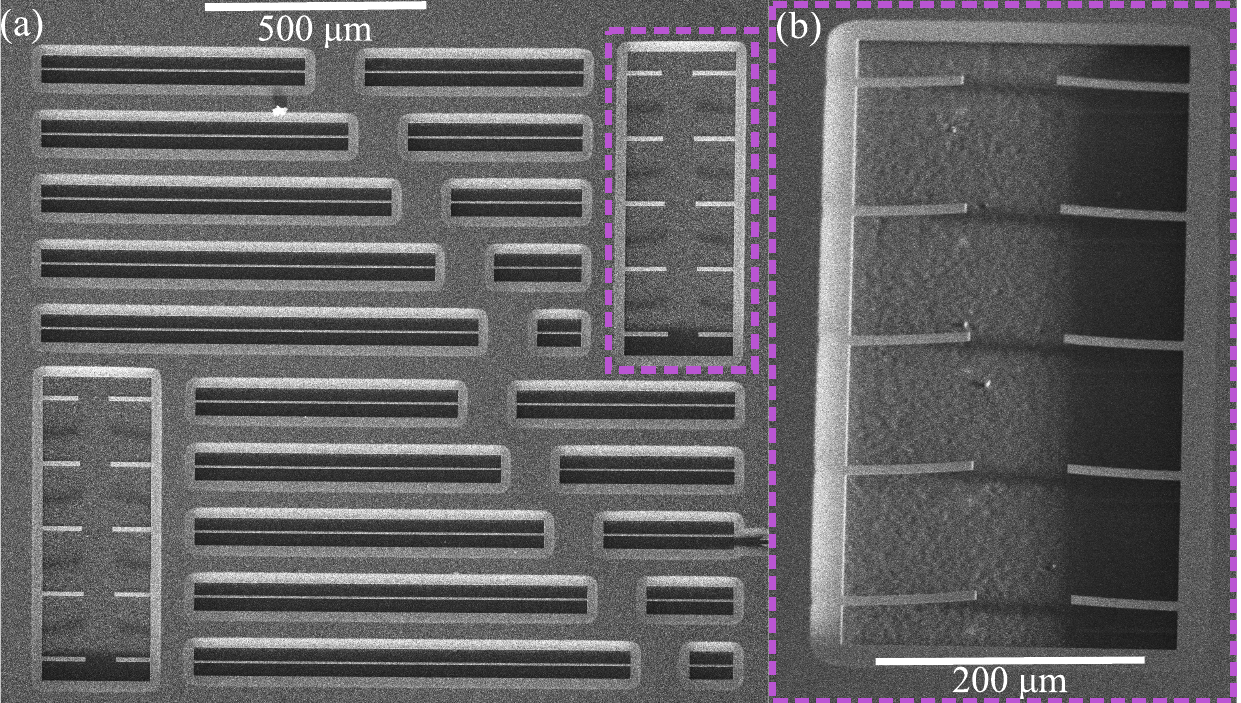}
\caption{(a) SEM image of arrays of uniform nanomechanical string and cantilevers used for material characterization. Width of devices is kept constant to $5\mu \textnormal{m}$, whereas the lengths of both cantilevers and strings are varied. The purple dashed box encompasses the array of cantilevers highlighted in (b). (b) Higher magnification SEM image of nanomechanical cantilevers with different lengths.}\label{fig:StanSEMs}
\end{figure*}

\subsection{Nanostrings}

\begin{figure}
\centering
\includegraphics[width=0.48\textwidth]{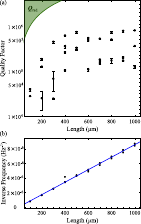}
\caption{(a) Quality factor versus length of uniform string nanomechanical resonators. Here each point represents an individual device with error bounded by the standard deviation of many ringdown measurements. The green curve represents the quality factor limit associated with radiation loss ($Q_{\textnormal{rad}}$) for devices with largest damping due to this loss (thickest). (b) Inverse resonance frequency as a function of string resonator length, where each point represents an individual device's fundamental transversal resonance frequency and the blue line represents the expected resonance frequency using the analytical model in Equation~\ref{eq:eigenfreq}, and extracted mean stress of the devices (172~MPa).}\label{fig:QvsLStringsChar}
\end{figure}

The additional set of uniform strings is fabricated on the same sample as the cantilever devices, as shown in Figure~\ref{fig:StanSEMs}(a). In this device set, the lengths and thicknesses of the string resonators vary, and the device widths are constant. Within this uniform string sample set we measure 30 string resonators of consistent $5~\mu\textnormal{m}$ width, whose lengths vary between $100~\mu \textnormal{m}$ and $1000~\mu\textnormal{m}$, and thicknesses span between $400~\textnormal{nm}$ and $815~\textnormal{nm}$. We plot the average and standard deviation of the quality factor of the fundamental transversal mode of each device as a function of length in Figure~\ref{fig:QvsLStringsChar}(a). The measured quality factors, as high as $Q=8.0\cdot10^5$, exceed those found in cantilever resonators due to the presence of tensile stress and therefore dissipation dilution. The resonance frequency of a string resonator of length and material density, $L$ and $\rho$, is related to the tensile stress, $\sigma$ by~\cite{s_schmid_fundamentals_2023} 

\begin{equation}\label{eq:eigenfreq}
f=\frac{{n}}{2L}\sqrt{\frac{\sigma}{\rho}}.
\end{equation}

\noindent Thus, the intrinsic tensile stress of each device can be determined by its resonance frequency. To determine the mean intrinsic tensile stress of the entire data set, we plot the inverse of the resonance frequency against length in Figure~\ref{fig:QvsLStringsChar}(b). As expected from Equation~\ref{eq:eigenfreq}, the data lies on a line. A fit then provides the mean stress of roughly $\sigma=172~\textnormal{MPa}$ after release.

 In contrast to the cantilever resonators, we find that the string resonators exhibit non-negligible radiation loss. The radiation loss of the fundamental mode of a string resonator follows~\cite{s_romero_engineering_2020} 
\begin{equation}\label{eq:Qclamp}
Q_{\textnormal{rad}}=\alpha \frac{{3 \rho_{s}}}{2\rho} \sqrt{\frac{E_{s} \rho}{2 \sigma \rho_s}} \frac{L}{h}.
\end{equation}
In this model, $\rho_s$ and $E_s$ represent the density ($2650~ \textnormal{kg}/\textnormal{m}^3$) and Young's Modulus ($170~\textnormal{GPa}$) of the silicon substrate. We determine the fitting parameter $\alpha=317$ based on a least-square fits of the intrinsic quality factor as a function of alpha ($Q_{\textnormal{int}}(h,\alpha)$). The green curve in Figure~\ref{fig:QvsLStringsChar}(a) shows the radiation loss quality factor limit for the thickest devices. This represents the highest radiation loss among the ``low-aspect-ratio strings'' data set, and the shaded region beyond this curve encompasses the radiation loss expected for thinner devices. It is apparent from Figure~\ref{fig:QvsLStringsChar}(a) that the devices are not at the radiation loss limit but the radiation loss is non-negligible. For example, for the $400~\mu \textnormal{m}$ long string with the highest quality factor ($Q=6.5\cdot 10^5$), radiation loss accounts for approximately $8.5\%$ of the total loss. As such, radiation loss was included in the intrinsic damping model of 4H-SiC, discussed in the main text.

\clearpage
\section{Material Limited Devices}

\begin{figure}[ht]
\centering
\includegraphics[width=0.48\textwidth]{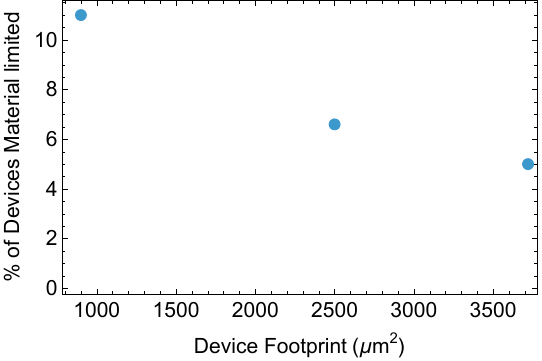}
\caption[Matlimit]{Percentage of material limited devices among a resonator sample set as a function of device footprint (device area). The sample sets represent cantilevers, low-aspect-ratio nanostrings and high-aspect-ratio nanostrings from smallest to highest footprint respectively.}\label{fig:MaterialLimFootprint}
\end{figure}

To gain further insight into the statistics of fabricated devices which reach the material limit we conduct analysis on the three different resonator geometries studied in this work. This includes 20 high-aspect-ratio nanostrings, 30 low-aspect-ratio nanostrings and 18 cantilever resonators. We seek to determine the dependency of quantity of material limited devices as a function of the device's total footprint. To do so, we calculate the number of devices which reach the volumetric damping material limit for each resonator geometry. We then find the percentage of material limited devices compared to the total number of devices measured for each geometry. Lastly, we plot this versus the device footprint which is fixed for high-aspect-ratio nanostrings and cantilevers, and an average for the low-aspect ratio nanostrings. The results are shown in Figure~\ref{fig:MaterialLimFootprint}, where we observe lower rates of material limited devices as we increase device footprint. This supports the idea that our thin films have a uniform density of defects, where some resonators possess no defects and reach the material limit. Given this, we would expect the trend observed which implies that increasing the size of devices increases the likelihood of encountering material defects.

\section{Overhang Effect Towards Dissipation Dilution Factor of String Resonators}\label{sec:AppUndercutAffect}

As can be seen in SEM figures within the main text, our devices possess overhang near the clamping points due to undercut during the fabrication process. Therefore they are not entirely perfectly clamped beams and to conduct a thorough analysis of intrinsic quality factors we summarize the effect of overhang towards the dissipation dilution factor here. We omit the analysis for cantilevers due to the lack of tensile stress (one clamping point). We infer the undercut to be 25 microns based on SEM images, and is fixed for each string regardless of the strings dimensions.

\begin{figure}
\centering
\includegraphics[width=0.48\textwidth]{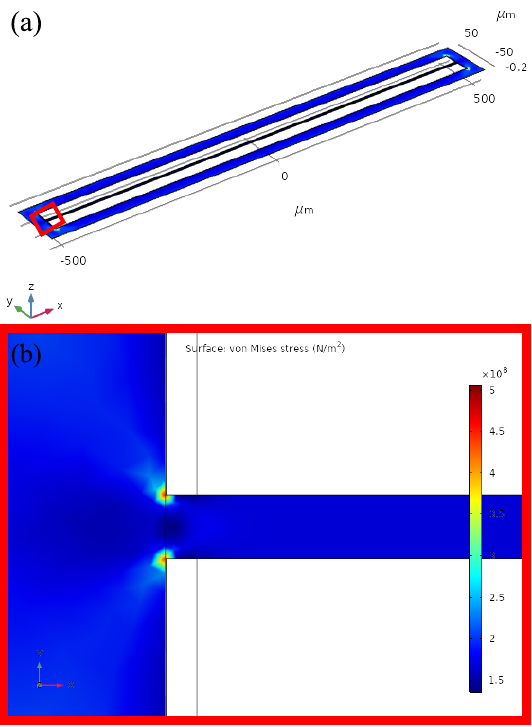}
\caption[COMSOL Overhang Stress]{(a) Simulated geometry of uniform string resonators with a rectangular frame, as closely representing actual device release from undercut etching. (b) Stationary solution of one clamping point of string resonator under tensile prestress of 220~MPa. }\label{fig:OverhangCOMSOl}
\end{figure}

First we model the stationary solution of our uniform strings of many different lengths since the proportion of undercut to total string length is different across our entire sample set. This can be seen in Figure~\ref{fig:OverhangCOMSOl}, where the deposition (unrelaxed) stress is 220~MPa. In many approximations the relaxed tensile stress is not determined by the thickness~\cite{s_fedorov_generalized_2019,s_ghadimi_elastic_2018}. This was confirmed via FEM for the sake of checking accuracy of the model and this assumption. As expected we find the fixed overhang makes a larger difference in the relaxed stress of shorter strings, specifically a $14\%$ difference for strings with overhang compared to perfectly clamped strings. The relaxed stress becomes very similar to the model of a perfectly clamped uniform string at longer lengths (within $4\%$ at 1mm length). Overall, we find the presence of overhang increases the tensile stress of the devices which supports others work~\cite{s_yao_relaxation_2022,s_buckle_universal_2021}.

We then calculate the eigenfrequency of the fundamental transverse mode for the entire range of lengths. Since the frequency of a stressed string scales as the square root of stress the overhang should make a smaller impact in resonance frequency~\cite{s_schmid_fundamentals_2023}. Indeed, we find that the resonance frequencies of strings with overhang to be very similar to those without overhang. Previous work studying the effect of overhang on the frequency of resonators has also shown similar results~\cite{s_verbridge_macroscopic_2007}.

\begin{figure}[h]
\centering
\includegraphics[width=0.48\textwidth]{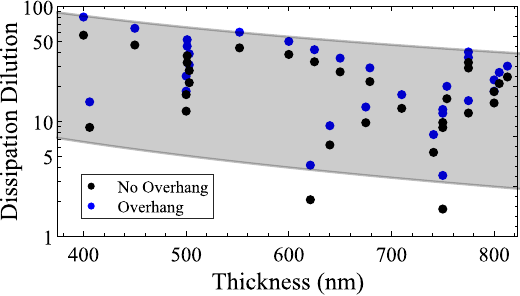}
\caption[Effect of Overhang on Dilution Factor]{Change in dissipation dilution factor for strings which have no overhang (black) and strings with overhang (blue). The gray bounded region represents the range where we'd expect the dissipation dilution values to be given the data set's mean stress, and lengths of the shortest and longest resonators in the data set.}\label{fig:DDUndercut}
\end{figure}

Following this, we then calculate the dissipation dilution factor for our mean string thickness across a variety of lengths that cover our entire data set. We reach a similar conclusion that strings with larger undercut relative to overall length are most affected. As a whole we observed increased dissipation dilution for the simulated overhang geometry in Figure~\ref{fig:OverhangCOMSOl}(a). We attribute this to the localization of relaxed stress near the strings clamping points as shown in Figure~\ref{fig:OverhangCOMSOl}(b). We then consider overhang for the entire range of thicknesses and lengths used for these devices and calculate the corrected dissipation dilution value for each device size. We plot these results in Figure~\ref{fig:DDUndercut}, directly comparing expected dissipation dilution values for uniform strings with (blue) and without (black) overhang. For comparison, we bound the data within an expected dissipation dilution value range (gray shaded region) using the analytical solution for a uniform width string resonator~\cite{s_schmid_fundamentals_2023}. We determine this expected range for our resonators using the data set's mean stress, as well as the shortest and longest resonator length as a function of thickness. We observe that accounting for overhang allows us to encompass every data point within the expected range of dilution values. For the approximations of intrinsic Q in the main text, we use the dissipation dilution terms accounting for overhang. 

\newpage

{
\makeatletter
\renewcommand{\@biblabel}[1]{[S-#1]}
\renewcommand{\bibnumfmt}[1]{[S-#1]}
\makeatother


}

\end{document}